\documentclass[a4paper]{spie}  
\usepackage[utf8x]{inputenc} 
\usepackage[]{graphicx}
\usepackage{subfigure}
\usepackage{transparent}
\usepackage{amssymb}
\usepackage{amsmath} 
\usepackage[english]{babel}
\usepackage{multicol}

\newcommand{\e}[1]{\times 10^{#1}} 
\newcommand{\ml}[1]{\mathrm{#1}} 

\newcommand{\specialcell}[2][l]{%
  \begin{tabular}[#1]{@{}l@{}}#2\end{tabular}}
\newcommand{\bit}{\begin{itemize}}
\newcommand{\bits}{\begin{itemize}\addtolength{\itemsep}{-0.5\baselineskip}} 
\newcommand{\ei}{\end{itemize}}

\title{The latest results from DICE (Detector Interferometric Calibration Experiment)} 

\author{A. Crouzier\supit{1}, F. Malbet \supit{1}, F. Henault\supit{1}, A. Léger\supit{3}, C. Cara\supit{2}, J. M. LeDuigou\supit{4}, O. Preis\supit{1}, P. Kern\supit{1}, A. Delboulbe\supit{1}, G. Martin\supit{1}, P. Feautrier\supit{1}, E. Stadler\supit{1}, S. Lafrasse\supit{1}, S. Rochat\supit{1}, C. Ketchazo\supit{2}, M. Donati\supit{2}, E. Doumayrou\supit{2}, P. O. Lagage\supit{2}, M. Shao \supit{5}, R. Goullioud\supit{5}, B. Nemati\supit{5}, C. Zhai\supit{5}, E. Behar\supit{1}, S. Potin\supit{1}, M. Saint-Pe\supit{1}, J. Dupont\supit{1}
\skiplinehalf
\supit{a}Institut d'Astrophysique et de Planétologie de Grenoble, 414 Rue de la Piscine, St Martin d'Hères, Grenoble, France; \\
\supit{b}Commissariat à l'Energie Atomique et aux Energies Alternatives, Saclay, centre d'études nucléaires de Saclay, Paris, France; \\
\supit{c}Institut d'Astrophysique Spatiale, Centre universitaire d'Orsay, Paris, France; \\
\supit{d}Centre National d'Etudes Statiales, 2 place Maurice Quentin, Paris, France; \\
\supit{e}Jet Propulsion Laboratory, 4800 Oak Grove Drive, Pasadena, CA, U.S.A. 91109
}

\begin{document} 
\maketitle 
\graphicspath{ {./Figures/} }
\begin{abstract}
Theia is an astrometric mission proposed to ESA in 2014 for which one of the scientific objectives is detecting Earth-like exoplanets in the habitable zone of nearby solar-type stars. This objective requires the capability to measure stellar centroids at the precision of $1\e{−5}$ pixel. Current state-of-the-art methods for centroid estimation have reached a precision of about $3\e{−5}$ pixel at two times Nyquist sampling, this was shown at the JPL by the VESTA experiment. A metrology system was used to calibrate intra and inter pixel quantum efficiency variations in order to correct pixelation errors. The Theia consortium is operating a testbed in vacuum in order to achieve $1\e{−5}$ pixel precision for the centroid estimation. The goal is to provide a proof of concept for the precision requirement of the Theia spacecraft. 

The testbed consists of two main sub-systems. The first one produces pseudo stars: a blackbody source is fed into a large core fiber and lights-up a pinhole mask in the object plane, which is imaged by a mirror on the CCD. The second sub-system is the metrology, it projects young fringes on the CCD. The fringes are created by two single mode fibers facing the CCD and fixed on the mirror. In this paper we present the latest experiments conducted and the results obtained after a series of upgrades on the testbed was completed. The calibration system yielded the pixel positions to an accuracy estimated at $4\e{-4}$ pixel. After including the pixel position information, an astrometric accuracy of $6\e{-5}$ pixel was obtained, for a PSF motion over more than 5 pixels. In the static mode (small jitter motion of less than $1\e{-3}$ pixel), a photon noise limited precision of $3\e{-5}$ pixel was reached.
\end{abstract}




\section{INTRODUCTION}\label{sec:INTRODUCTION} 

\subsection{Astrometric exoplanet detection method}\label{Astrometric exoplanet detection method}

With the present state of exoplanet detection techniques, none of the rocky planets of the Solar System would be detected. By measuring the reflex motion of planets on their central host stars, astrometry can yield the mass of planets and their orbits. However it is necessary to go to space to reach the precision required to detect all planets down to the Earth mass.

A European consortium has been proposed several versions of astrometric missions capable of detecting Earth-like planets in the habitable zone of nearby stars, in the framework of M ESA missions. The first version proposed was NEAT\cite{Malbet11,Malbet12,Malbet13,2014SPIE.9143E..2LM}, which used formation flying, the second version is Theia\cite{MalbetSPIE16}, which is a more standard single spacecraft TMA telescope. Both concepts use differential astrometry to detect and measure the mass of planets down to the level of an Earth mass in the habitable zone of nearby stars. We want to explore in a systematic manner all solar-type stars (FGK spectral type) up to at least 10 pc from the Sun. For both mission a critical component of the telescope is the metrology system which projects dynamic Young fringes on the detector plane. The fringes allow a very precise calibration of the CCD in order to reach micro-pixel centroiding errors\cite{Crouzier14}.

One of the fundamental aspects of the NEAT mission is the extremely high precision required to detect exo-Earths in habitable zone by astrometry. The amplitude of the astrometric signal that a planet leaves on its host star is given by the following formula:
\begin{equation}\label{eq:astrometric_signal}
A = 3 \mu \ml{as} \times \frac{M_{\ml{Planet}}}{M_{\ml{Earth}}} \times \left(\frac{M_{\ml{Star}}}{M_{\ml{Sun}}}\right)^{-1} \times \frac{R}{1\ml{AU}} \times \left(\frac{D}{1\ml{pc}}\right)^{-1}
\end{equation}

Where $D$ is the distance between the sun and the observed star, $M_{\ml{Planet}}$ is the exoplanet mass, $R$ is the exoplanet semi major axis and $M_{\ml{Star}}$ is the mass of the observed host star. For an Earth in the habitable zone located at 10 pc from the sun, the astrometric signal is 0.3 micro arcseconds (or $1.45\e{-11}$ rad). This corresponds to a calibration of pixelation errors to $1\e{-5}$ for the Theia mission 

\subsection{The NEAT-demo / DICE testbed}\label{subsec:Presentation of the DICE}

\begin{figure}[t]
\begin{center}
\includegraphics[height = 80mm]{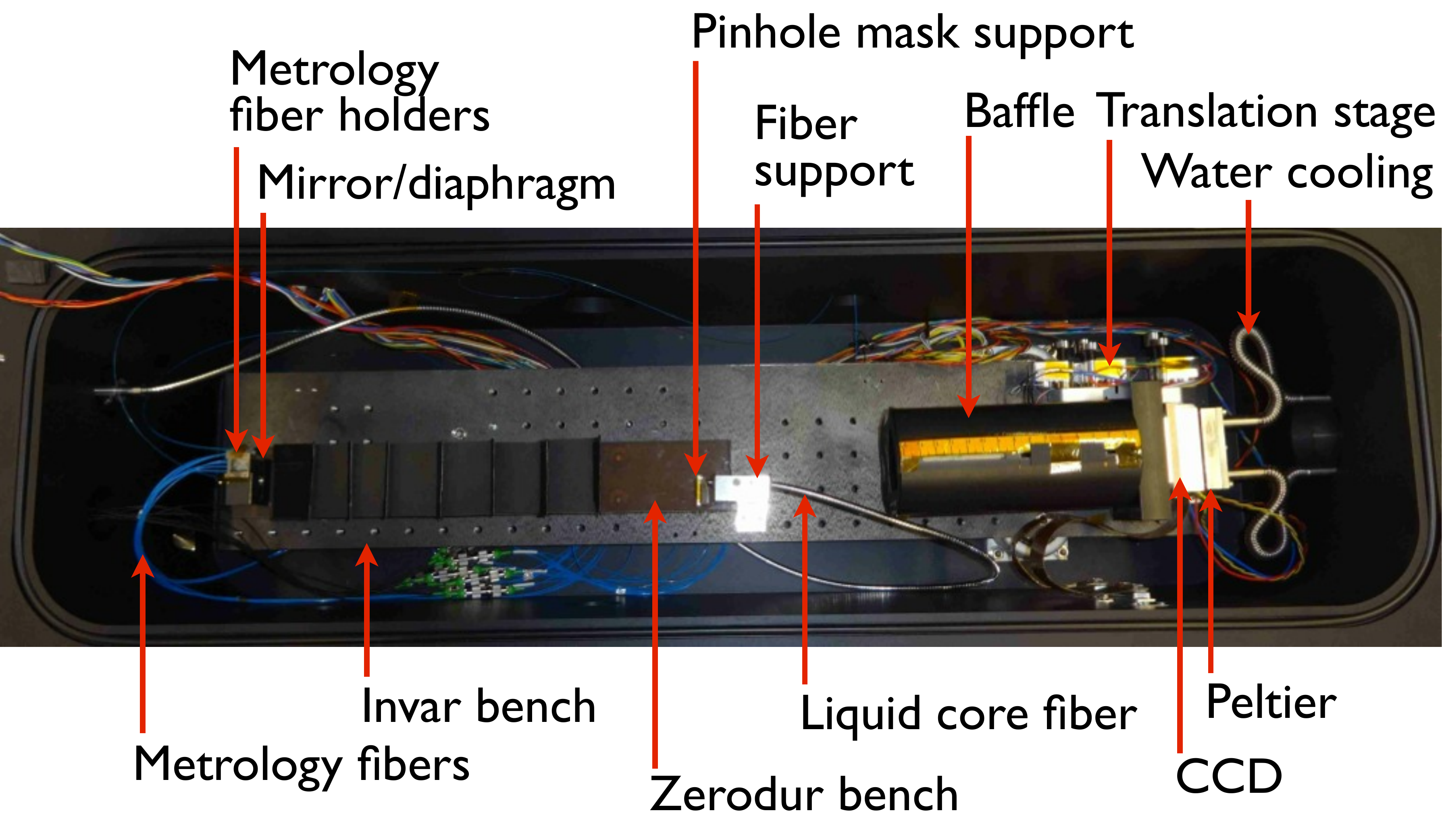}
\caption{\label{schematic_vacuum_chamber}\textbf{Schematic of the optical bench of the NEAT testbed.} The part shown here is the bench inside the vacuum chamber, which mimics the optical configuration of the NEAT spacecraft: it is the core of the experiment. Two main peripheral sub-systems, the metrology and pseudo stellar sources produce respectively fringes and stars on the CCD.}
\end{center}
\end{figure}

In order to demonstrate the feasibility of the calibration, a testbed called NEAT-demo was assembled at IPAG. Figure \ref{schematic_vacuum_chamber} is a labeled picture of the inside of the vacuum chamber. Here we will not elaborate any longer about the testbed, in previous SPIE papers we have already presented: the testbed itself, the context around it, its specifications, a photometric budget (Amsterdam, 2012)\cite{Crouzier12}{}, the results after the first light was obtained in July 2013 and an error budget (San Diego, 2013)\cite{Crouzier13}{}. A last paper presents a more complete and updated error budget (Montreal, 2014)\cite{2014SPIE.9150E..0IH}. In the context of Theia the testbed has been renamed DICE (Detector Interferometric Calibration Experiments), we will use that name in this paper. 

Since the last published results\cite{Crouzier14}, significant upgrades on the testbed have been completed. The CCD was swapped with the 1 quadrant (1 amplifier) version, instead of 4 quadrant (4 amplifiers). This allowed to solve the electronics ghosting issue which was impairing the astrometric measurements. Additionally, the light baffle was improved in many aspect, the result is a lower level of stray light and an improvement of the metrology accuracy. The new experimental results obtained after the upgrades are presented in the next section.

\subsection{EXPERIMENTAL RESULTS}

\subsubsection{Dark and flat-fields}\label{xpdata:dark_and_flat}

\begin{figure*}[t]
\centering
\subfigure[Mean dark frame]{\label{darkXpData}
\includegraphics[width=0.49\textwidth]{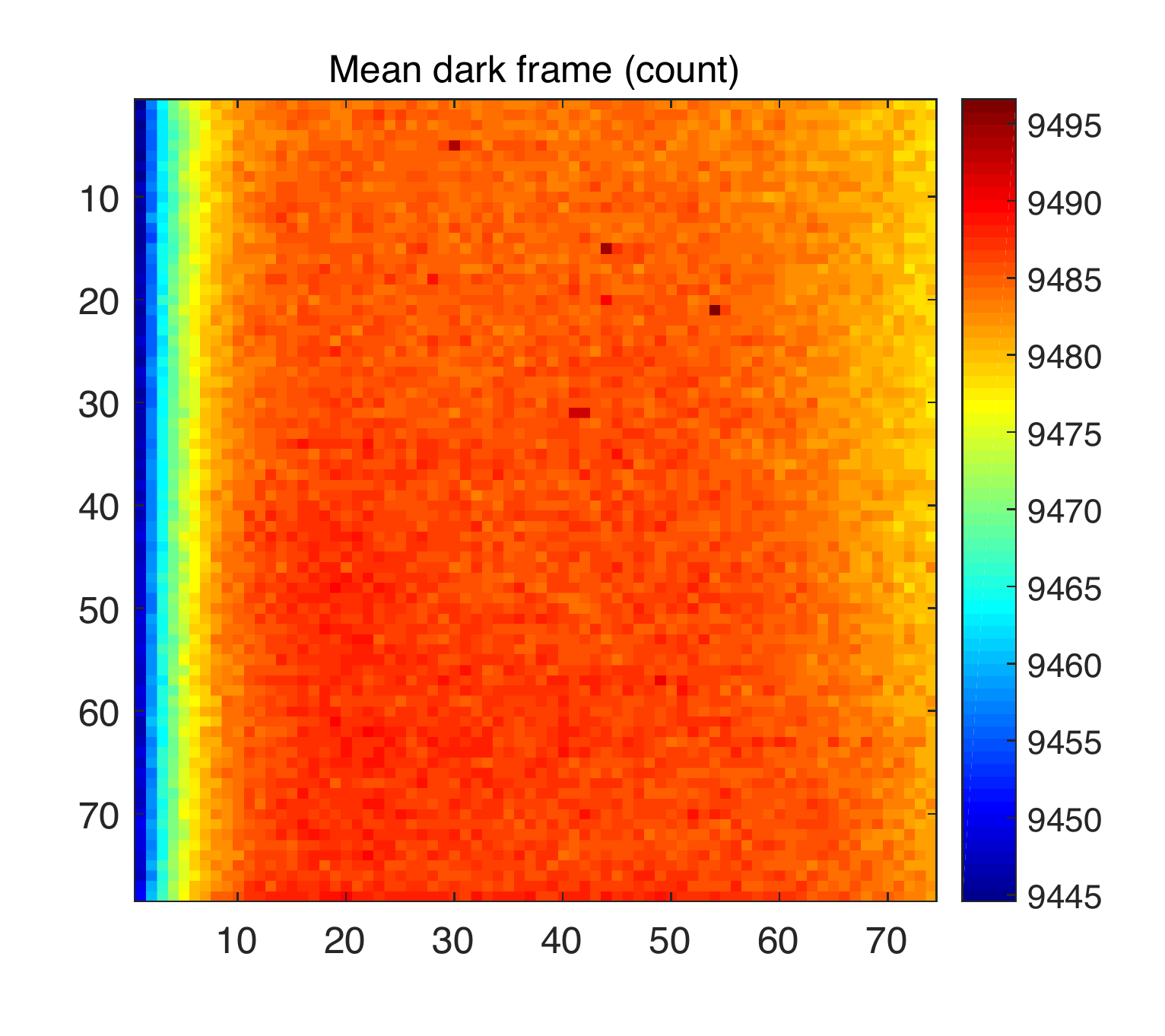}}
\subfigure[Temporal noise in dark data cube]{\label{darkNoise}
\includegraphics[width=0.475\textwidth]{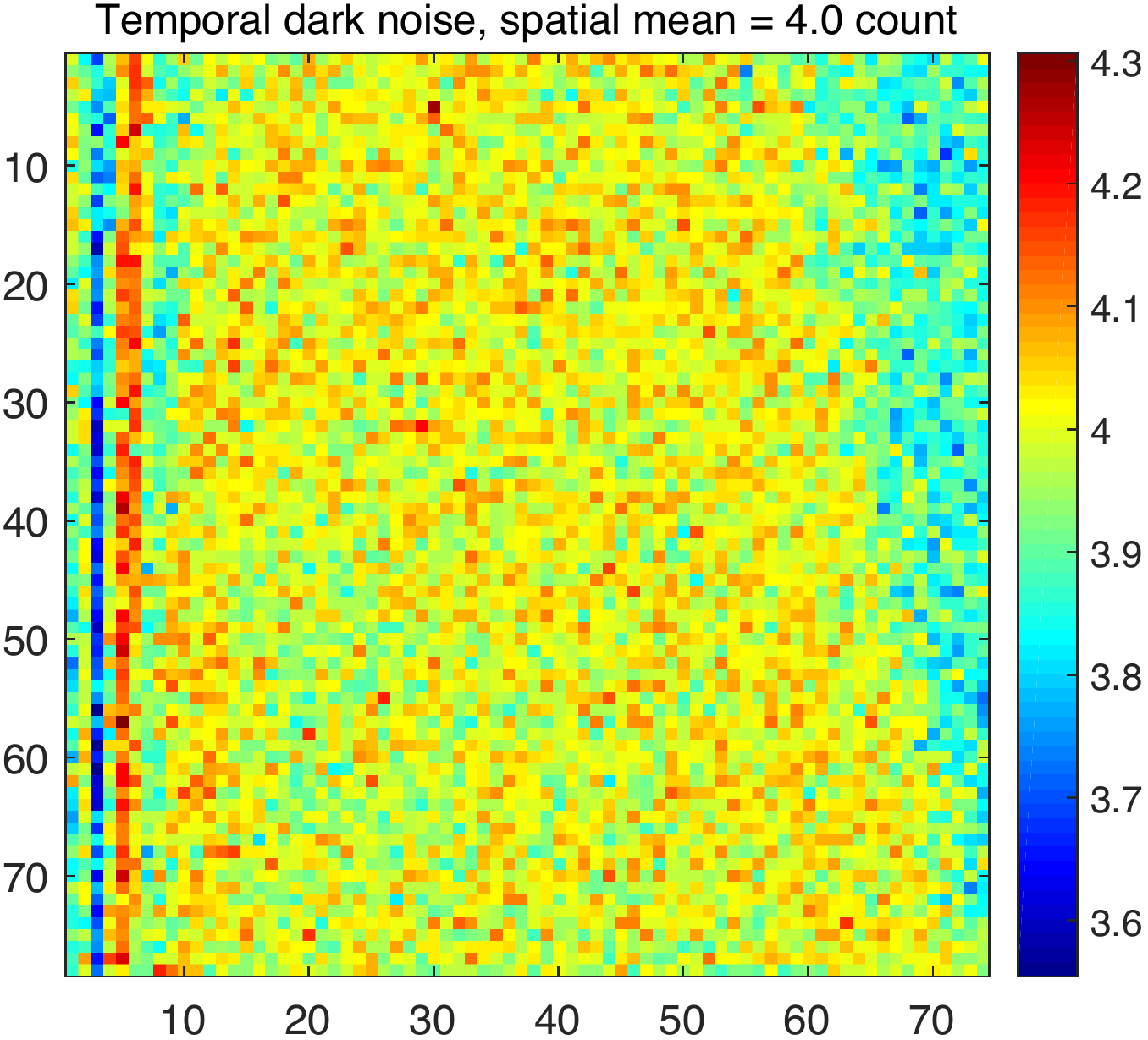}}
\caption{\label{darkCube}Temporal mean and noise of dark data cube}
\end{figure*}

Figure \ref{darkXpData} shows a mean dark frame. Figure \ref{darkNoise} shows the temporal standard deviation of the data cube which yielded said mean dark frame. The low level of dark+readout noise (4.0 counts SD) is small compared to the photon noise: the CCD is operated at the significant fraction of full well, typical photon noises are 30 to 100 counts.

\begin{figure*}[t]
\centering
\subfigure[Measured PRNU with a white LED and multimode fiber]{\label{prnu}
\includegraphics[width=0.485\textwidth]{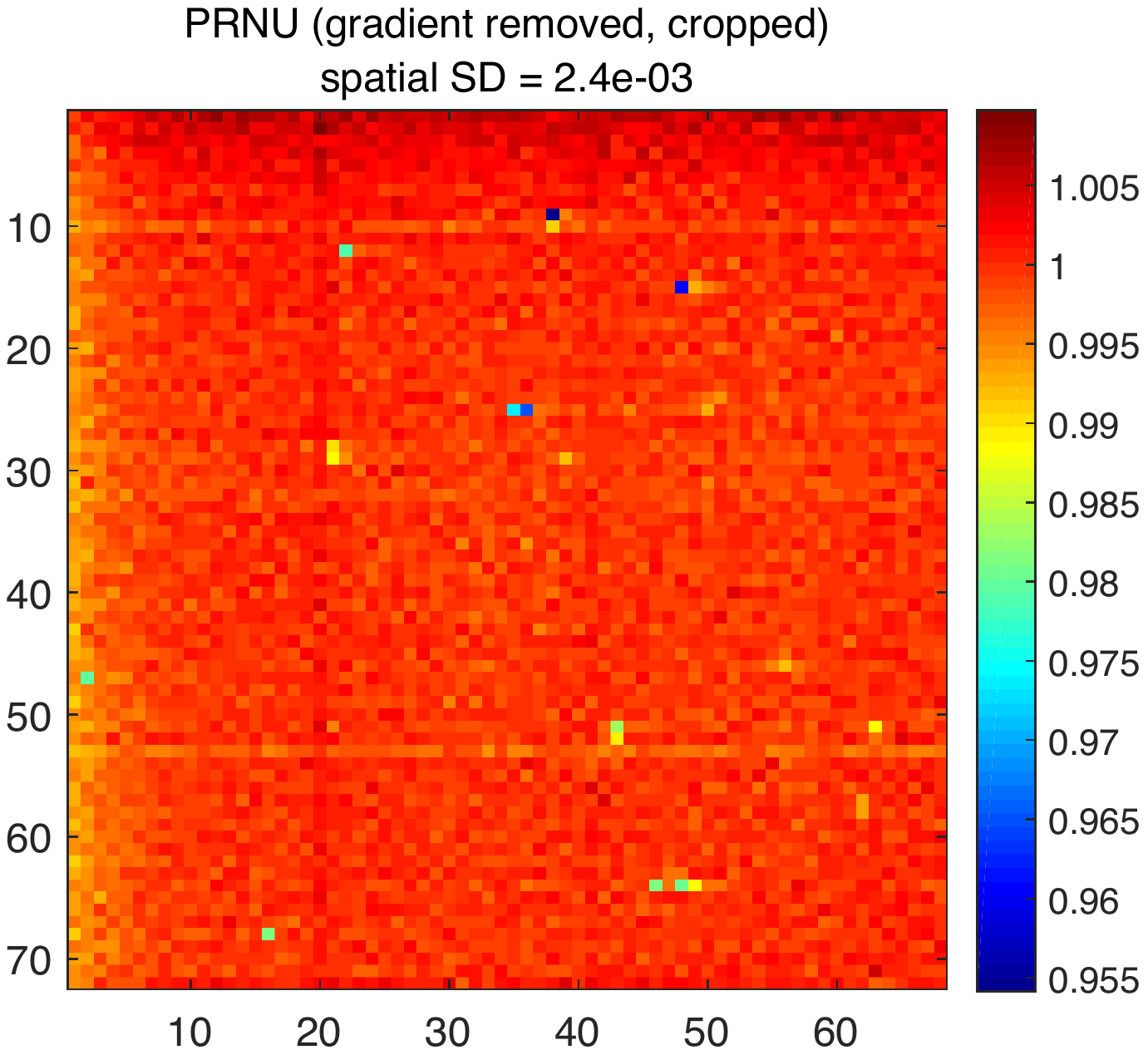}}
\subfigure[Difference between two measured PRNU, with a displacement of the fiber tip (photon noise limit: $2\e{-5}$)]{\label{prnuDiff}
\includegraphics[width=0.48\textwidth]{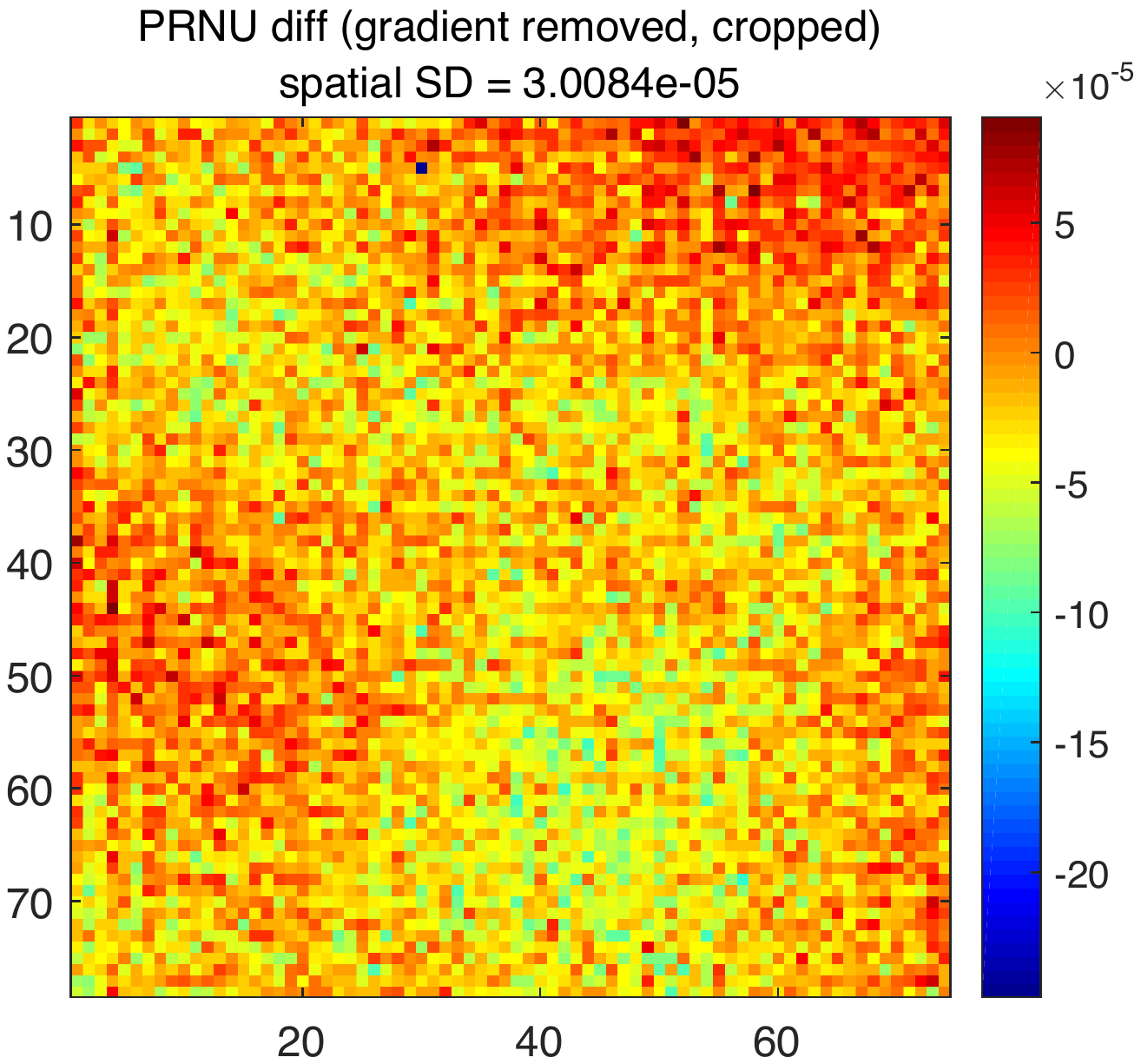}}
\caption{\label{mainLabel}PRNU and its difference with a displacement of the fiber tip.}
\end{figure*}

Figure \ref{prnu} shows the measured PRNU, after processing of the flat field obtained with the white LED and multimode fiber. The measured PRNU RSD is $2.4\e{-3}$. The PRNU distribution is clearly not pure white noise, several features can be distinguished: 2 horizontal bright bands, dark pixels down to 95\% efficiency (caused by dust contamination) and a background looking roughly like white noise but with also faint low frequency patterns and more visible biases close to the upper and left edges.

In order to estimate the systematics on this result, the PRNU difference between two different fibers position was computed (Fig. \ref{prnuDiff}). The fiber motion (about 2 cm) is significant, even compared to the size of the baffle field of view, yet the magnitude of the difference is small: $3\e{-5}$, very close to the photon noise limit at $2\e{-5}$. Additionally, applying the Allan deviation method on the flat field data cube did not reveal any anomalies. This suggests that the PNRU map obtained is of high quality. Using flat fields from the metrology was attempted. But the data was clear, PRNU differences showed RSD about $1\e{-3}$ to $5\e{-4}$ (depending on the residual level of stray light), but in all cases significantly higher than for white flats.

There was an additional concern about the possibility of having systematic effects, and in particular speckles, caused by the large core (365 $\mu$m diameter) of the multimode fiber. To estimate these effects an experiment similar to the one described above was performed: a flat field difference after variation of the fiber bending (instead of a fiber displacement) was measured. It showed the same amplitude difference of about $3\e{-5}$, ruling out a significant impact from the bending.

\subsubsection{Metrology}

\begin{figure*}[t]
\centering
\subfigure[Horizontal pixel offsets]{\label{xOffsetsXpData}
\includegraphics[width=0.48\textwidth]{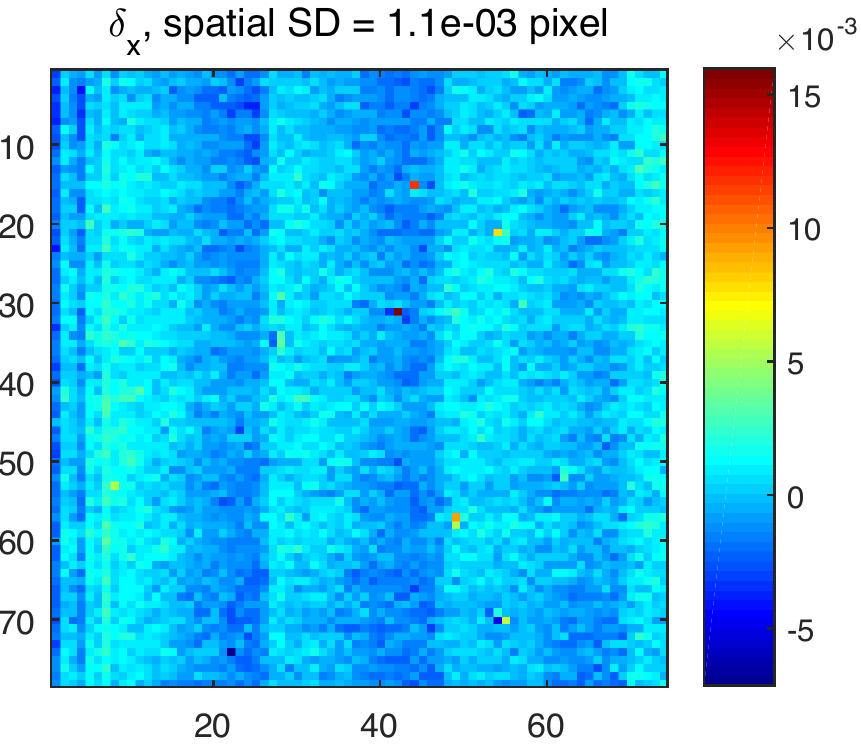}}
\subfigure[Vertical pixel offsets]{\label{yOffsetsXpData}
\includegraphics[width=0.48\textwidth]{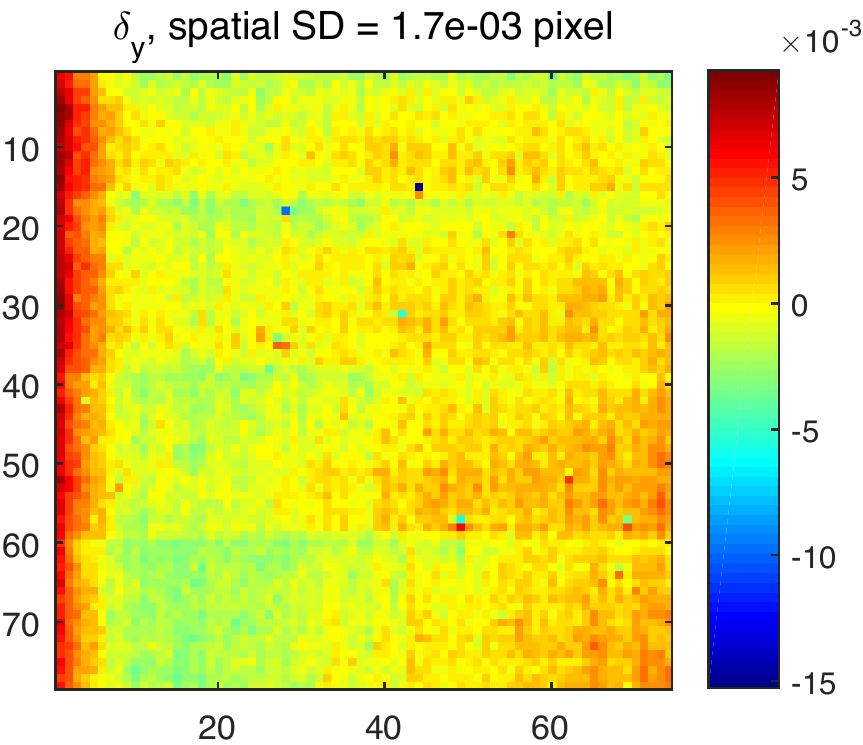}}
\caption{\label{mainLabel} Pixel offsets (in pixel units) as measured by the metrology (baseline H1H4/V1V4).}
\end{figure*}

One complete data set consists of a pair of baselines (vertical and horizontal). The results presented are with the laser coherence control system active (coherence $\approx$ 1 cm). Figures \ref{xOffsetsXpData} and \ref{yOffsetsXpData} show the measured pixel offsets for baseline H1H4/V1V4, in respectively the horizontal and vertical directions.

\begin{figure*}[t]
\begin{center}
\includegraphics[width = \textwidth]{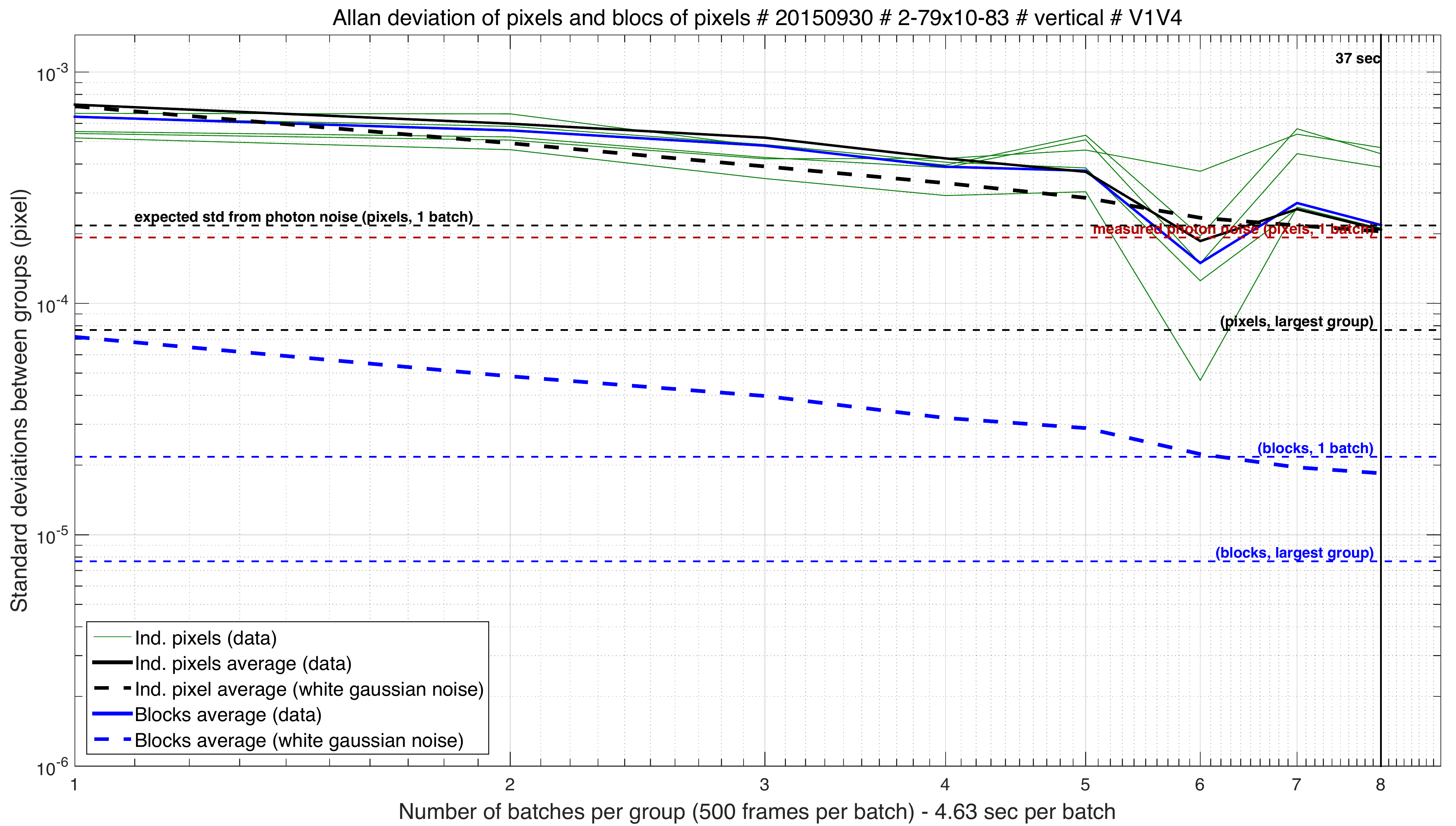}
\caption{\label{allanDevXpData} Allan deviations of projected pixel offsets (vertical baseline).}
\end{center}
\end{figure*}

Figure \ref{allanDevXpData} shows the Allan deviation of projected pixel offsets for the vertical baseline. The final deviation for individual pixels is $2\e{-4}$ pixel. In spite of all the measures used against stray light (coherence reduction and baffling), there is no improvement. The precision is far from the photon limit and correlated noise is still present. The result for the horizontal baseline is very similar (same final deviation, same issues with correlated noise).

\begin{figure*}[t]
\centering
\subfigure[Horizontal pixel offsets difference]{\label{xOffsetsDiff}
\includegraphics[width=0.48\textwidth]{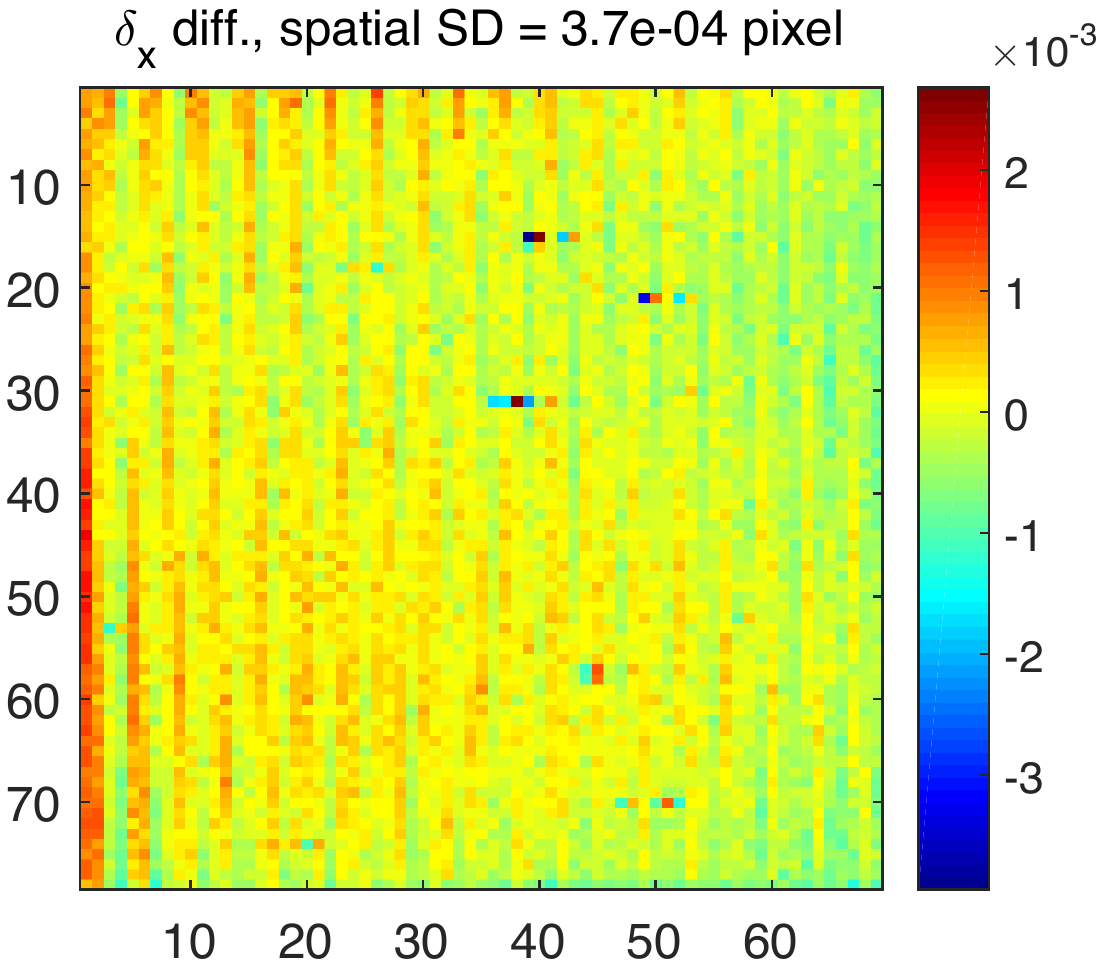}}
\subfigure[Vertical pixel offsets difference]{\label{yOffsetsDiff}
\includegraphics[width=0.48\textwidth]{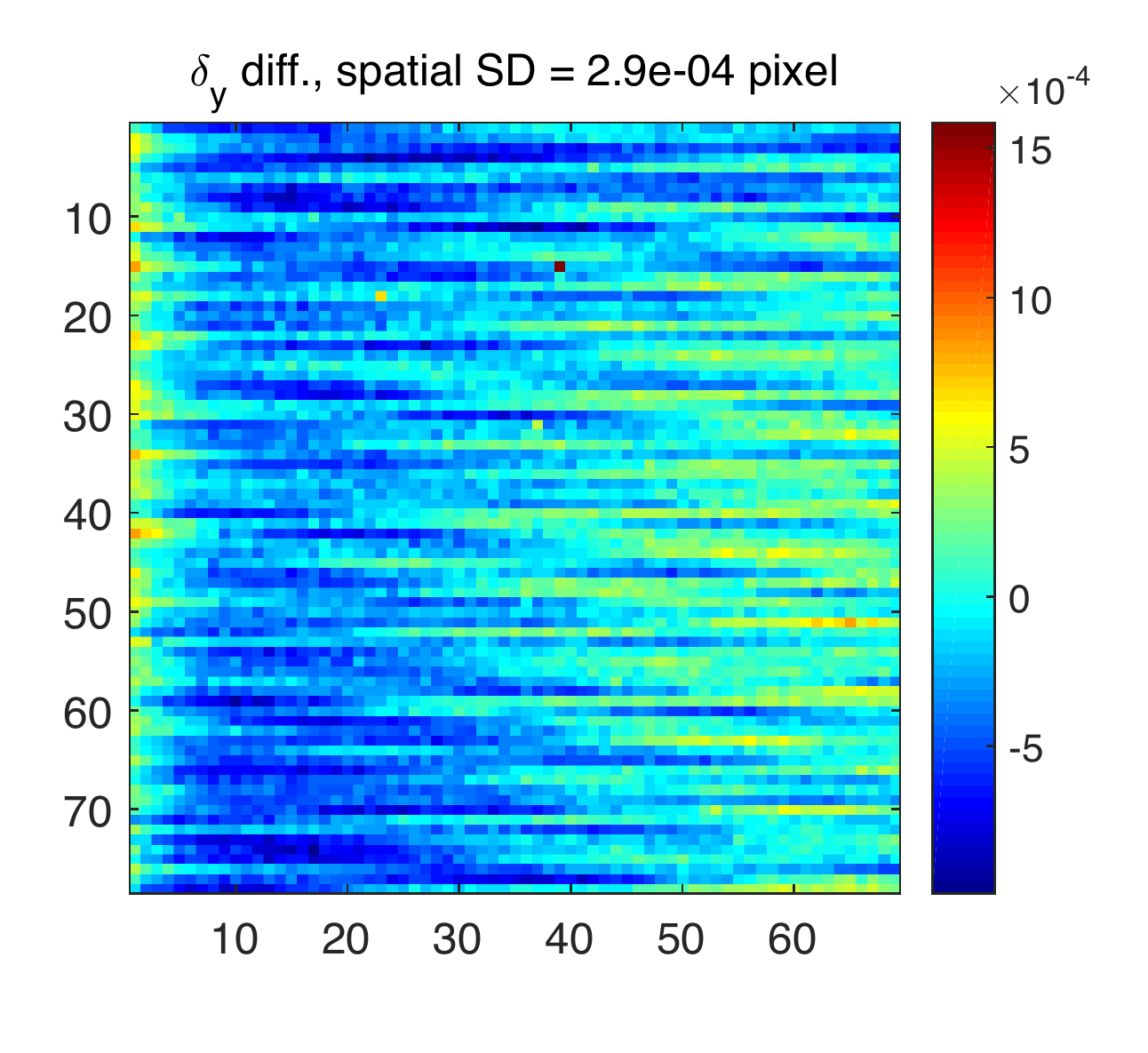}}
\caption{\label{offsetsDiff}Pixel offsets differences, between two different sets of metrology baselines.}
\end{figure*}

A more rigorous way to estimate the accuracy of pixel offsets measurements is to perform two independent analysis for two physically different pairs of baselines and to compare the results. Each pair have different separations, producing respectively fringes spacings of 2.4 and 4 pixels on the detector, and are located at slightly different places in the vacuum chamber (the angles of incidence of metrology beams are changed). Because of this, at least the stray light bias produced in each configuration will be different. Figures \ref{xOffsetsDiff} and \ref{yOffsetsDiff} shows the difference between results for baselines H1H4/V1V4 and baselines H2H4/V2V4, for respectively horizontal and vertical pixel offsets.

From both the pixel offsets (Fig. \ref{xOffsetsXpData} and \ref{yOffsetsXpData}) and their differences (Fig. \ref{xOffsetsDiff} and \ref{yOffsetsDiff}) it is clear that the dust contamination has biased the pixel offset measurements to at least several parts per thousand. However the effect is very localized (one or a few pixels) and should have a small effect on astrometric measurements perhaps easily identifiable at it should affect centroids independently and at precise and known locations.

The amplitudes of the differences have standard deviations up to about $4\e{-4}$ pixels, which is slightly larger than the value given by Allan deviations. This indicates that static systematics (constant in time) are present in the data. In this case the visible systematics depend of the physical position of the metrology fibers. Moreover, this map of the difference presents some structures and high frequency components which look like speckles. This strongly suggests that stray light is still an issue, but other unknown systematics are possible as well. This data only gives a lower bound for the amplitude of systematics in the measured pixel offsets. The ultimate indication that the metrology measurements are correct is that the addition of pixel offsets into the data pipeline should improve the astrometric accuracy (see next Section).

\subsubsection{Pseudo stars}\label{sec:xp data pseudo stars}

The pseudo star data which yields the best results is the one presented here (taken in ambient air). The translation stage supporting the detector was moved into 90 different positions by small steps of about 0.17 pixel each\footnote{0.17$\pm\;0.01$ pixel} into an roughly straight, almost vertical line. To each detector position corresponds a data cube of pseudo star data (the detector is not moved during acquisition). The \emph{single-position} analysis has not revealed any problem with the data. The precision obtained is about $6\e{-5}$ pixel for all detector positions, when splitting each data cube into 4 batches. This corresponds to the value expected for theoretical photon noise. Extrapolating the photon noise to the whole cubes yields an expected photon noise accuracy limit at $3\e{-5}$ pixel.

\begin{figure}[t]
\begin{center}
\includegraphics[width = 0.5\textwidth]{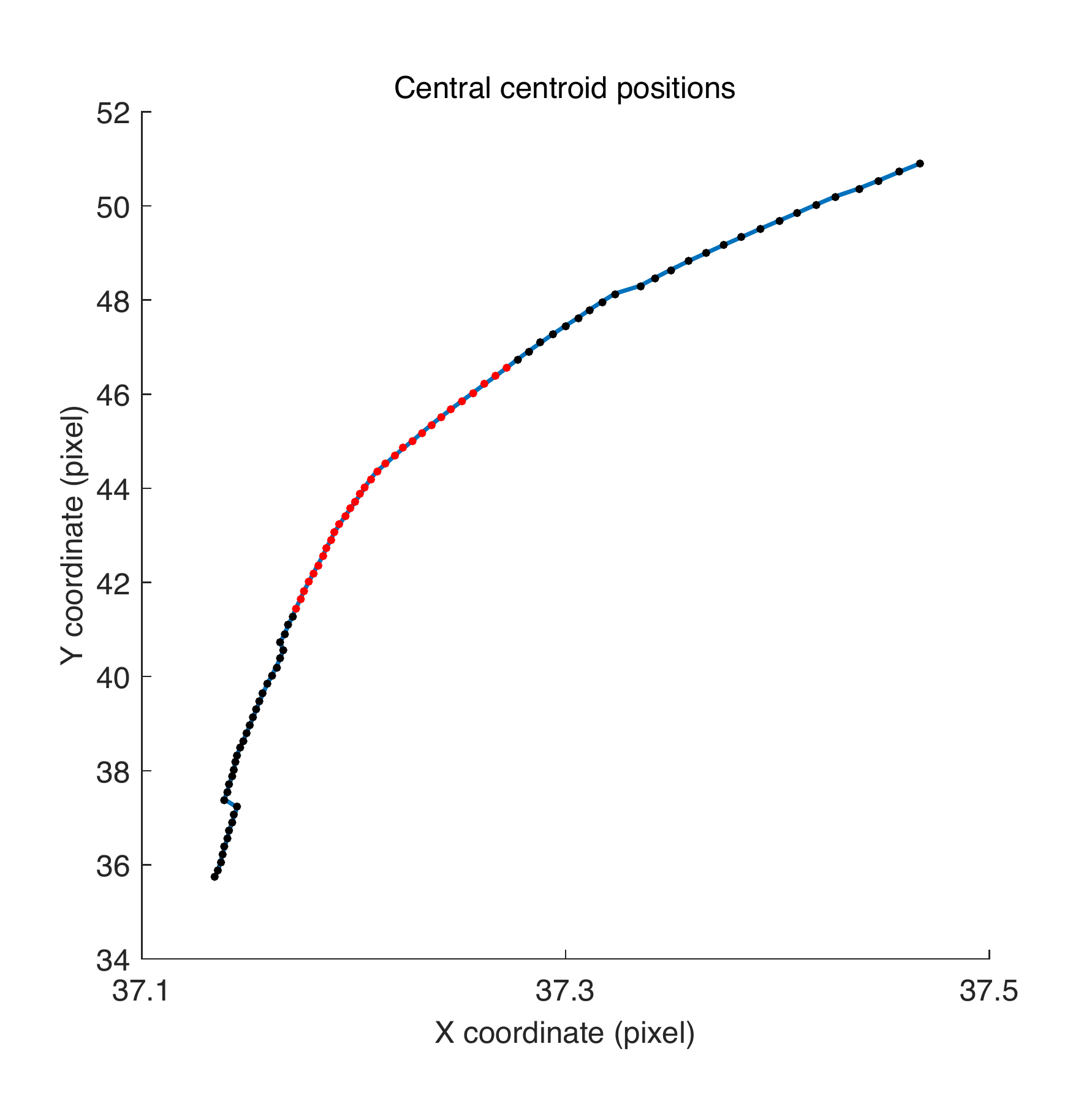}
\caption{\label{ccdPositionsAutocorr} Central centroid coordinates for the 26/09/2015 pseudo star dataset. Note that the scale is much smaller for the horizontal axis (the real line is almost vertical ). Indexes between 25 to 57 (final dataset for accuracy) are in red.}
\end{center}
\end{figure}

The cause of the sloppy detector motion (roughly straight, almost vertical line) is the erratic behavior of the translation stage which uses piezoelectric motors in an open loop. Figure \ref{ccdPositionsAutocorr} shows the pixel coordinates of the central centroid from positions 1 to 90.  The commands sent to the stage for each step were identical and for a purely vertical motion. 

The distortion of the translation has no major consequence for the analysis. However there is a much more significant issue with the translation stage, which is intrinsic to its design. It uses piezoelectric motors (against springs) as actuators. To produce the translation, two parallel sliders are pushed by one motor each. This design allows for rotation motion by using opposite directions with the sliders. But for translation, as each motor pushes in an irregular and unequal manner, this also creates significant unwanted tip tilt. When discussing the astrometric accuracy, this crucial point will be developed. 

\begin{figure*}[t]
\begin{center}
\includegraphics[width = 0.85\textwidth]{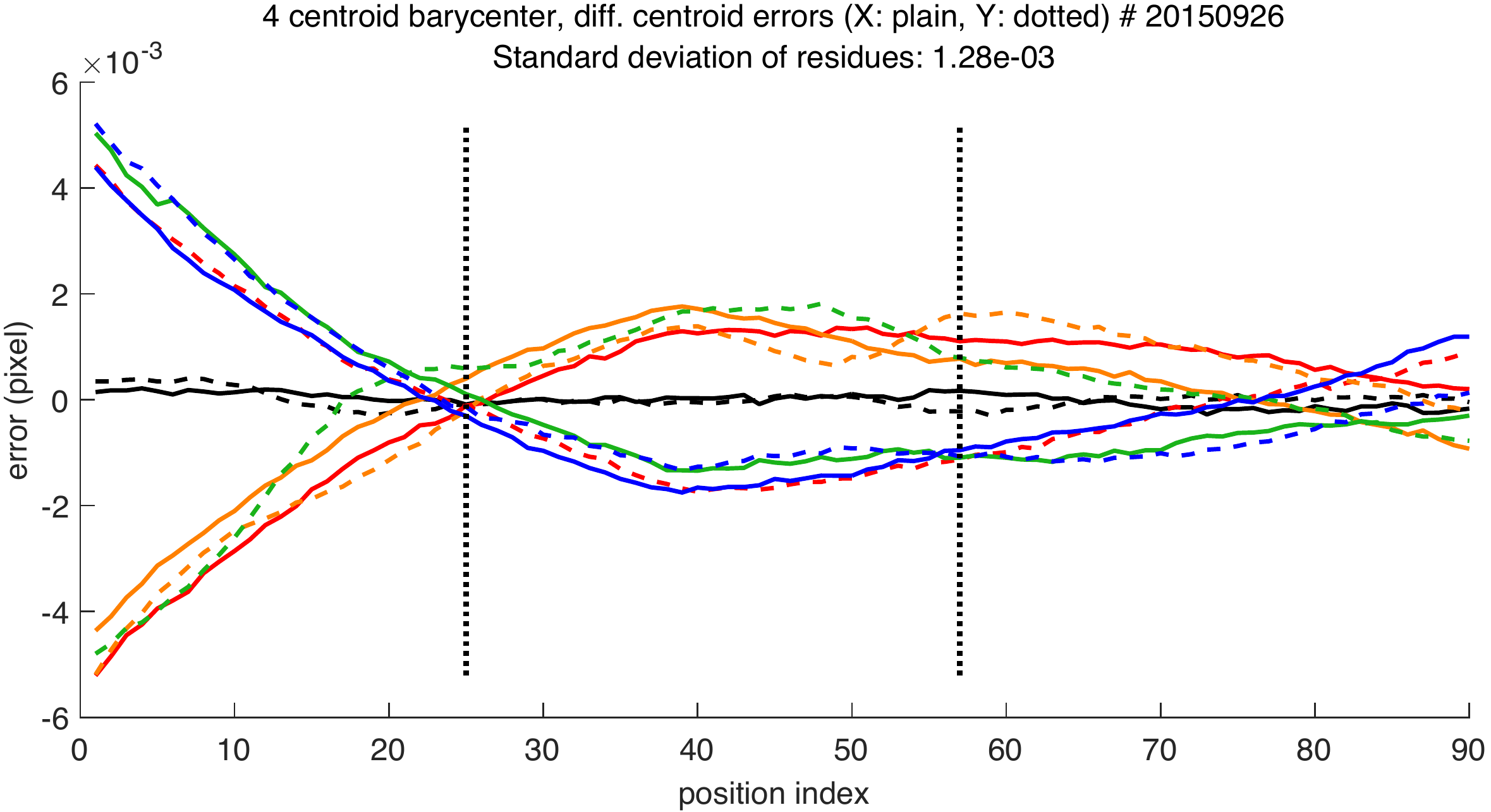}
\caption{\label{20150926_barycenter_residuals_pos1to90} Barycenter residuals for each position (from index 1 to 90). Each centroid has a different color (black centroid is the central one). Plain lines are X axis residuals, dotted lines are Y axis residuals. Indexes between 25 to 57 (final dataset for accuracy) are indicated by the vertical dotted lines.}
\end{center}
\end{figure*}

\begin{figure}[t]
\begin{center}
\includegraphics[width = 0.5\textwidth]{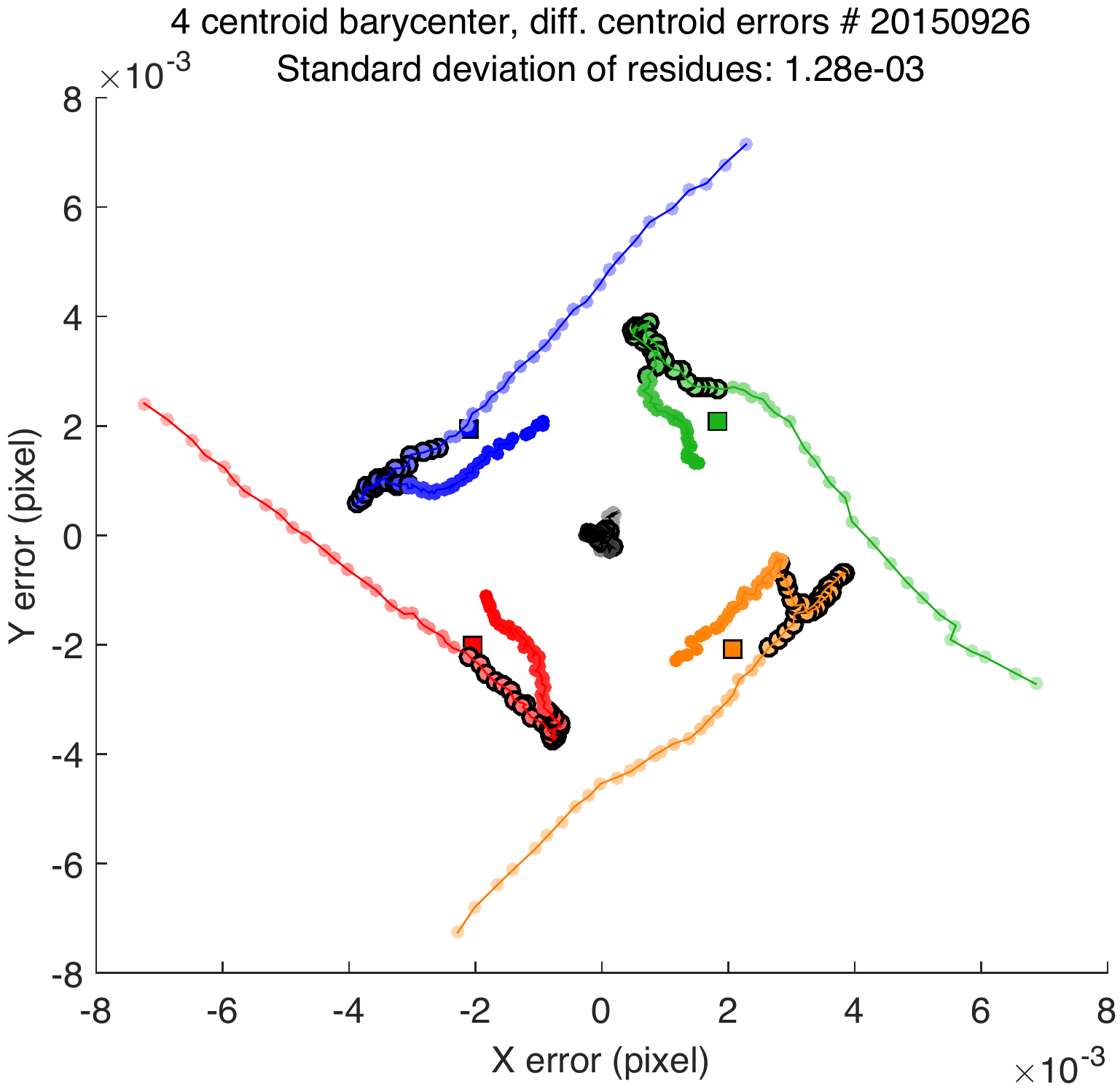}
\caption{\label{20150926_barycenter_residualsXvY_pos1to90} Barycenter residuals for each position (from index 1 to 90). Each centroid has a different color (black centroid is the central one). The relative positions between pseudo stars (illustrated by the squares) have been downscaled to correspond to the magnitude of the residuals. Indexes between 25 to 57 (final dataset for accuracy) are circled in black.}
\end{center}
\end{figure}

Figure \ref{20150926_barycenter_residuals_pos1to90} shows the residuals obtained from with a simple barycenter method, for all positions (1 to 90). More precisely, the residuals are the variations of the projected distances (horizontal and vertical axis) between each star and the barycenter of all the other ones. The large amplitude drifts of residuals are produced by a corresponding drift of the translation stage tip/tilt axis as a function of the position index. Correlations/anti-correlations between the stars are based on the geometric layout, as show by Fig. \ref{20150926_barycenter_residualsXvY_pos1to90}, where the same residuals have been plotted in a parametric manner (X versys Y). The unwanted motion is mainly a roll, but some tip/tilt is present as well.

\begin{figure*}[t]
\begin{center}
\includegraphics[width = \textwidth]{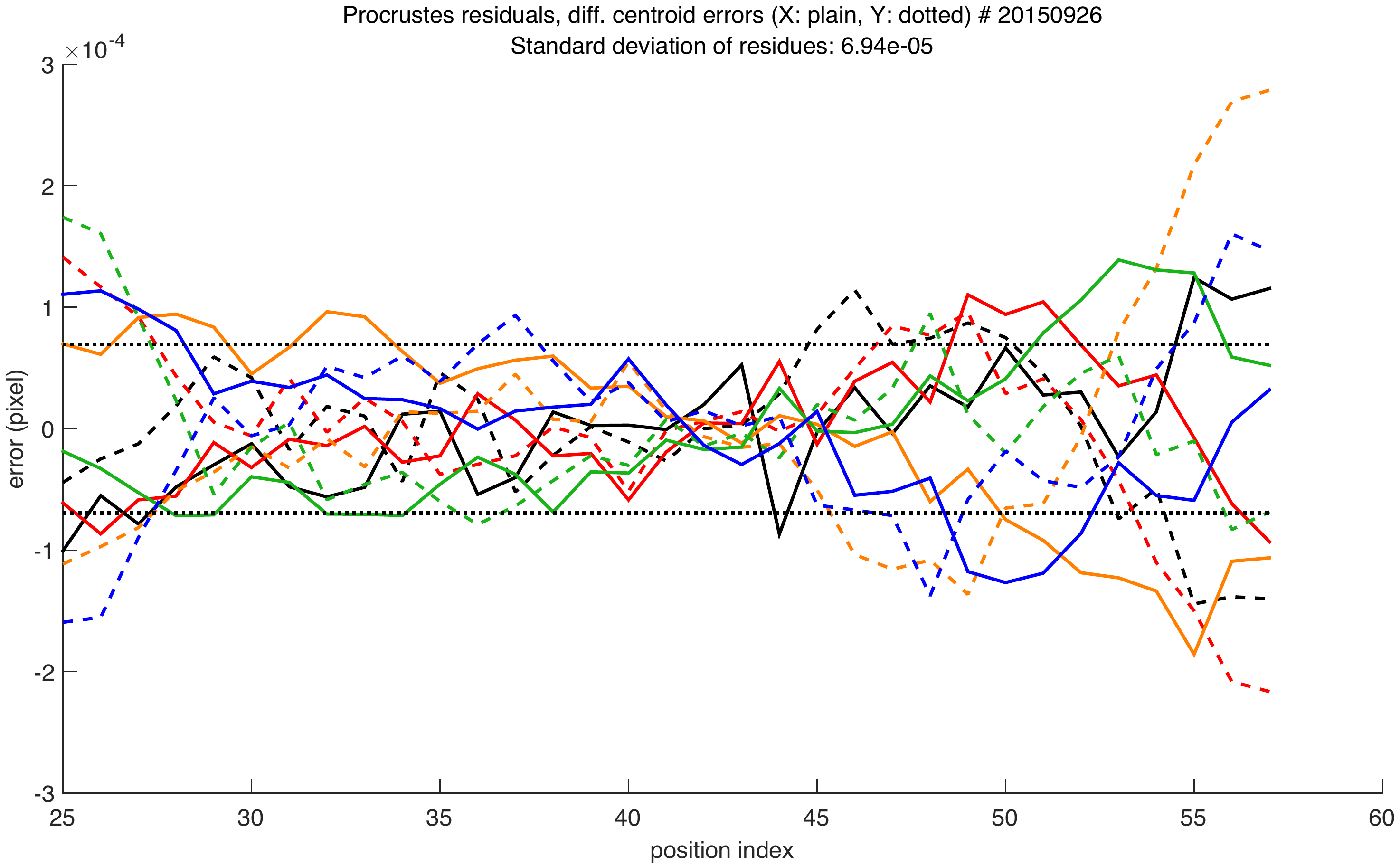}
\caption{\label{20150926_procrustes_residuals_pos25to57} Procrustes residuals for each position (from index 25 to 57). In this case the PRNU and pixel offset calibration are used. Each centroid has a different color (black centroid is the central one). The horizontal dotted lines indicate the dispersion of the residues ($\pm \sigma$).}
\end{center}
\end{figure*}

To minimize the impact of this issue, an area with relatively small tip tilt errors was selected, corresponding to positions 25 to 57, spanning 5.4 pixels. Figure \ref{20150926_procrustes_residuals_pos25to57} shows the Procrustes residuals for this range of indexes. The function of the Procrustes method is precisely to compensate for the geometrical effects, after PRNU and pixel offsets corrections. As expected, the residuals are much smaller than with the simple barycenter technique.

Table \ref{tab:Accuracies with different types of calibrations} shows the final accuracies (after Procrustes residuals) for the 25-57 position indexes, in 4 different calibration setups. The PRNU calibration (from flat field) and the pixel offset calibration (from metrology fringes) can each be activated or not. The final accuracy corresponding to each case is given by the table. These accuracies are the result of a \textit{multi-position} analysis, in this case the motion is quasi-linear over a length of 5 pixels.

A test of spatial averaging over several detector positions was done. Instead of considering the residuals from each position separately, the centroids are averaged over groups of positions. The accuracy is then given by the Procrustes residuals over the centroids corresponding to each group. The 33 positions are binned into groups in the following way: positions 1 to 11, positions 12 to 22, positions 23 to 33, effectively forming 3 juxtaposed segments (without interlacing). In table \ref{tab:Accuracies with different types of calibrations}, accuracies corresponding to individual \emph{positions} and \emph{groups} of positions are respectively flagged by (p) and (g). Note that for the groups, the photon limit ($3\e{-5}$ for individual positions) goes below $1\e{-5}$.

\begin{table}[t]
\caption{Accuracies with different types of calibrations (in pixel units).}\label{tab:Accuracies with different types of calibrations}
\begin{center}
\begin{tabular}{c c c}
\hline
- & No flat (pixel) & flat  (pixel)\\
\hline
\hline
No metrology & \specialcell{(p) $2.1\pm0.12\e{-4}$\\(g) $1.4\pm0.27\e{-4}$} & \specialcell{(p) $1.7\pm0.10\e{-4}$\\(g) $1.6\pm0.30\e{-4}$} \\
\hline
Metrology & \specialcell{(p) $2.9\pm0.17\e{-4}$\\(g) $2.1\pm0.40\e{-4}$} & \specialcell{(p) $9.7\pm0.56\e{-5}$\\(g) $5.9\pm1.1\e{-5}$}  \\
\hline

\multicolumn{3}{l}{\specialcell{The (p)~/~(g) flags respectively indicate accuracies over\\positions~/~groups of positions.}}
\end{tabular}
\end{center}
\end{table}

The best result was obtained when the flat field and metrology calibration were combined, obtaining a final astrometric accuracy of $9.7\pm0.56\e{-5}$ for individual positions and $5.9\pm1.1\e{-5}$ when binning into 3 groups. The error bars displayed on the table are from Monte Carlo simulations with random astrometric jitter, independent for each star, as is expected from photon plus pixelation noise. When considering residual calibration noise, the error bars are good upper limits: if other uncalibrated systematics are present they can only increase the final noise.

\section{DISCUSSION}

\subsection{Bias on pixel offsets}

There are several clues indicating that stray light is a significant if not the main source of errors for the pixel offset measurement:
\bit
\item flat field differences with fiber tip motion in coherent light show high bias: speckles with a relative contrast up to $1\e{-3}$ are visible in this case.
\item when comparing the pixel offsets obtained with two different pairs of baselines, there is a difference significantly above photon noise
\ei
The results on the flat fields seem to be partly in contradiction with the results on the pixel offsets. While tests have show that with the reduced coherence length there is a significant attenuation of the speckle contrast (about a factor 3), no gain was observed on the metrology with reduced coherence. One possibility is that diffraction and/or diffusion on the baffle vane edges is the main source of noise on pixel offsets. In this case the OPD between direct and stray light is around 1 cm plus or minus a factor 2 (depending on which vane is considered), so the contrast attenuation obtained is very small. Another attempt to improve the metrology was to average the results from 9 different fiber tip positions to provide angle diversity, but no gain was obtained. More theoretical work or simulations are needed to explain all these observations. 

The pixel offset difference is however about 3 times smaller than the pixel offset SD, so useful calibration information was obtained. The comparison of the astrometric accuracy with and without pixel offset calibration (after PRNU correction) has confirmed this: the accuracy was improved by about a factor two (when used with the flat field).

\subsection{Issue with tip/tilt error correction}

A critical issue was discovered with the Procrustes technique: it is in fact not powerful enough to correctly address the geometrical distortions down to the required accuracy. First, note that as each axis can change separately, plate scale changes can occur separately on the X and Y axis. The standard Procrustes technique only has one global scaling parameter, the technique used in this experiment is thus already a modified version with 2 independent scaling axis. But even that is not sufficient. As the stars are coming from different angles, the effect of tip/tilts are not strictly homogeneous plate scale changes. A geometrical analysis of the problem reveals that the plate scale inhomogeneity between two opposite stars is proportional to $\alpha\theta$, with $\alpha$ the angular separation between the stars and $\theta$ the tip/tilt amplitude. Fig. \ref{20150926_barycenter_residuals_pos1to90} and \ref{20150926_barycenter_residualsXvY_pos1to90} show these motions, in particular Fig.~\ref{20150926_barycenter_residualsXvY_pos1to90} very clearly shows a roll, to which the experiment is very sensitive and which is in principle not permitted by the translation stage. However this roll motion is not problematic since it is easily and cleanly subtracted. The tip-tilt is the real issue.

A more sophisticated way than Procrustes superimposition to inverse the projection was implemented. The new method uses 6 parameters to characterize the detector position: X, Y, Z positions, and tip, tilt and roll angles. From these parameters and centroid angular coordinates (2 per centroid) their positions on the detector can be obtained through exact projection in 3D. A gradient descent method is used to simultaneously minimize the centroid angular coordinates and the detector parameters for each data cube (one data cube per detector position). The number of parameters to fit is $10+6 N_{\ml{cubes}}$, for $10 N_{\ml{cubes}}$ data points. At this point in the development process the proper working of the deprojection algorithm is uncertain. Monte Carlo simulations seem to indicate that only 5 centroids are not enough to properly retrieve the deprojection without overfitting. The problem may be too poorly constrained with too many free parameters. With only random astrometric jitter, the centroid position residuals are correlated and grossly underestimated. 

In our setup, $\alpha=1.6\e{-3}$ rad (separation defined as the side of the square formed by the pseudo stars, which is 40 pixels). The translation stage tip-tilt amplitude is about $5\e{-3}$ rad (peak to valley), as measured with both Procrustes and the projection inversion technique. So astrometry systematic noise induced can go up to $3\e{-4}$ pixel in the worst case and between opposite stars.


\subsection{Effectiveness of calibrations}

Table \ref{tab:Accuracies with different types of calibrations} displays several interesting features:
\begin{enumerate}
\item Spatial averaging of the centroid always improves the accuracy, although the gains from one configuration to the next are rather unequal. This could be caused by the large error bars associated with this measurement.
\item PRNU calibration improves the accuracy for individual positions, and further improvement is obtained by adding the metrology calibration.
\item Using the metrology alone deteriorates the accuracy.
\end{enumerate}

Having on the same dataset observations 2 and 3 is very surprising, so far no convincing explanation has been found. On most previous data sets both the PRNU and metrology have had impacts on the astrometric accuracy of the same sign whether used alone or combined. Another startling result is the accuracy before PRNU, which is already very good. From a PRNU RSD measured at $2.4\e{-3}$, this was not expected. This result have been consistently obtained over all datasets. It seems that the good pixel to pixel homogeneity of the back-illuminated CCD chosen as a detector allow for a good baseline astrometric accuracy (without calibration). With the flat field only, a very interesting accuracy of $1.7\e{-4}$ is reached. A possible explanation for the good pre-flat accuracy is that the PRNU measured has a non Gaussian distribution (dust contamination, detector edge effects), it could thus have a different impact on astrometry than expected. The impact is minimized if the PSF stay clear of the dust and edges.

The low gain obtained from the flat field plus pixel offset calibrations is disappointing and intriguing. One possible reason for the low effectiveness is the spectral dependency of the pixel responses: the pixel offsets are measured at 633 nm whereas the pseudo stars cover much of the visible spectrum. CCDs, and in particular back-illuminated CCDs are known to show measurable spectral dependency \cite{1995PASP..107.1094R}. This spectral effect was not investigated in the experiment, the metrology system was designed for operation at 633 nm and the only source available was the HeNe laser.

\subsection{Spatial averaging}

The experimental results show that astrometric accuracy can be increased by spatial averaging, thus spreading the pixelation errors over more pixels than would be allowed by a single PSF. The gain is however very limited (e.g. from $9.7\e{-5}$ to $5.9\e{-5}$ pixel with all calibrations active). A much more useful data set for this technique would be a grid with a large number of detector positions. In fact an analysis was done over such a dataset (a grid of 340 positions, spacing step of 1 pix). A gain in accuracy of about a factor 30 (from $2.4\e{-3}$ to $7\e{-5}$ pixel) with groups of interlaced positions was obtained. With a spacing as small as 1 pixel, a good fraction of systematics cancels out in the differential astrometry, including in particular pixelation and tip tilt errors. Because of poor starting accuracy in this case (large tip tilt errors), the final result does not exceed the other result. It does however suggest a powerful way to mitigate pixelation systematics for a space mission.

\subsection{Other uses of the calibration}

The calibration was presented for the case of an astrometric measurement, but the data can also be analyzed under the angle of photometry, by looking at the relative intensities of the PSFs. Detection of the transit of an exo-Earth around a Sun requires a photometric stability better than $10^{-4}$, which can benefit from a detector calibration as well. The photometry is in fact available as a by product of the astrometric fit process. Although the testbed is not optimized for flux photometry, the relative photometric stability measured on the same dataset between two stars after calibration has reached $9\e{-5}$. 

\section{CONCLUSION}

The calibration system yielded the pixel positions to an accuracy estimated at $4\e{-4}$ pixel. After including the pixel position information, an astrometric accuracy of $6\e{-5}$ pixel was obtained, for a PSF motion over more than 5 pixels. Without the (flat and metrology) calibrations the astrometric accuracy is $1.4\e{-4}$ pixel  (all other things equal). With the \textit{single position} mode (small jitter motion of less than $1\e{-3}$ pixel), a photon noise limited precision of $3\e{-5}$ pixel was reached.

The \textit{single position} result shows that the detector and electronics dark and readout noises are well behaved and should not prevent reaching higher accuracies. The number that is relevant for an astrometric mission is the \textit{multi-position} analysis result: $6\e{-5}$ pixel. It characterizes the residual noise from pixelation errors after calibrations. As this accuracy was obtained for a motion over 5.4 pixels, a distance larger than the PSF diameter, it can be extrapolated to the whole CCD. In the DICE experiment the translation stage tip tilt is responsible for the larger errors associated with wider motions. The \textit{single position} results also confirm that turbulence (in the closed vacuum chamber) is not an issue. Some of the earlier datasets were taken in air and in vacuum but no accuracy gain was measured.

The main source of noise for the metrology seems to be stray light, despite numerous baffle upgrades targeted at the issue. More work is needed to confirm that there are no other important sources of systematics and to understand what are the best ways to mitigate stray light. The limited effectiveness of the metrology calibration on astrometric accuracy (2-3 fold improvement at best) could be caused by a spectral dependency of the PRFs. To first quantitatively assess and then mitigate the issue the same interferometric calibration should be performed at several wavelengths distributed across the visible spectrum.

The final objective was set at $5\e{-6}$ pixel. In reality, the exact requirement depends on the spacecraft parameters and scientific objectives. For the new mission concept Theia, it is only $10^{-5}$ pixel \cite{2015pathwaysMalbet,MalbetSPIE16}. Theia can still detect nearby habitable Earths, even if it is restricted to slightly fewer and closer stars. As the experimental data showed, it could be possible to significantly enhance the final accuracy by averaging the relative star coordinates over several detector positions. In a real mission, one can reasonably conceive that up to 100 different positions (per epoch) could be used, resulting in the best case into a relaxing of the specification up to a factor 10 (the maximum gain is given by the square root of the number of positions). However jittering is not desirable as a first approach to increase accuracy as it could impose significant additional constraints on the instrument capabilities, such as fast re-pointing, higher bandwidth or on-board processing (e.g. to "shift and add" images) and could decrease the overall instrument efficiency (e.g. the number of observable targets, or percentage of time spent collecting photons versus doing maneuvers).


\acknowledgments     

We would like to thank the engineering team at IPAG for their support. We acknowledge the labex OSUG@2020 and CNES for financing the experiment and CNES and Thales Alenia Space for funding the PhD of A.\ Crouzier. At last, we are grateful to Bijan Nemati, Chengxing Zhai and Inseob Hahn for hosting Antoine Crouzier into their team and taking the time to answer many of our questions. 


\bibliography{spie_article_antoine_crouzier_2016}   
\bibliographystyle{spiebib}   

\end{document}